\documentclass[english]{article}
\usepackage[T1]{fontenc}
\usepackage[latin9]{inputenc}
\usepackage{geometry}
\usepackage{color}
\usepackage{array}
\usepackage{graphicx}
\usepackage{times}

\usepackage{cite}
\makeatletter

\providecommand{\tabularnewline}{\\}

\makeatother

\usepackage{babel}
\begin{document}

\title{Quantum sealed-bid auction using a modified scheme for multiparty
circular quantum key agreement}

\author{Rishi Dutt Sharma$^{a,}$\thanks{Email: rishi.nitp@gmail.com}, Kishore
Thapliyal$^{b,}$\thanks{Email: tkishore36@yahoo.com}, Anirban Pathak$^{b,}$\thanks{Email: anirban.pathak@gmail.com}
\\
$^{a}$ National Institute of Technology Patna, Ashok Rajhpath, Patna,
Bihar 800005, India\\
$^{b}$ Jaypee Institute of Information Technology, A-10, Sector-62,
Noida, UP-201307}
\maketitle
\begin{abstract}
A feasible, secure and collusion-attack-free quantum sealed-bid auction
protocol is proposed using a modified scheme for multi-party circular
quantum key agreement. In the proposed protocol, the set of all ($n$)
bidders is grouped in to $l$ subsets (sub-circles) in such a way
that only the initiator (who prepares the quantum state to be distributed
for a particular round of communication and acts as the receiver in
that round) is a member of all the subsets (sub-circles) prepared
for a particular round, while any other bidder is part of only a single
subset. All $n$ bidders and auctioneer initiate one round of communication,
and each of them prepares $l$ copies of a $\left(r-1\right)$-partite
entangled state (one for each sub-circle), where $r=\frac{n}{l}+1$.
The efficiency and security\textcolor{blue}{{} }of the proposed protocol
are critically analyzed. It is shown that the proposed protocol is
free from the collusion attacks that are possible on the existing
schemes of quantum sealed-bid auction. Further, it is observed that
the security against collusion attack increases with the increase
in $l$, but that reduces the complexity (number of entangled qubits
in each entangled state) of the entangled states to be used and that
makes the scheme scalable and implementable with the available technologies.
The additional security and scalability is shown to arise due to the
use of a circular structure in place of a complete-graph or tree-type
structure used earlier.
\end{abstract}

\section{Introduction}

In our daily life, we often find it difficult to perform all the tasks
ourselves. Consequently, we outsource some tasks.  For outsourcing
a task at the lowest possible price (or to sell a product at the highest
possible price), we often use a process referred to as auction, which
is extremely relevant for imperfect market. Specifically, in an imperfect
market, it may be hard to find (a) an actual valuation of an item
which you wish to sell or buy, and (b) to identify potential buyers
or sellers. Auction helps us in obtaining these information. This
process (auction) is very important in today's society and is frequently
used in various forms. Interestingly, the notion of an auction is
almost as old as the civilization \cite{cassady1967auctions}. In
fact, this process links market and economics to cryptography and
thus to mathematics and computer science. Interestingly, the recent
developments in the field of quantum cryptography \cite{bennett1984quantum,ekert1991quantum,bennett1992quantum}
further extend this link and establish a link between physics (especially,
quantum mechanics) and market (see \cite{piotrowski2003quantum,patel2007quantum,piotrowski2008quantum,naseri2009secure}
and references therein). With time various types of auction mechanism
and the associated rules have been evolved. Based on those rules,
auction schemes may be classified into a few classes which includes
(but not restricted to), (i) English auction \cite{xiong2012bidder},
(ii) Dutch auction \cite{rockoff1995design}, and (iii) Sealed-bid
second-price (Vickrey) auction \cite{michiharu1998secure}. Here it
would be apt to note that in the English auction, one of the bidders
announces his bid publicly at the beginning of the auction process.
The competitor bidders announce higher price bids and the bid price
rises until we find there is no further bid. The bidder, who announced
the highest bid, wins the bid and pays his bid. In contrast, in a
Dutch auction, the auctioneer announces a high bid and then gradually
lowers the bid. The bidder who first accepts the bid, wins and he
has to pay that price. These bidding mechanisms are traditionally
performed publicly, where all bidders can listen the call of the other
bidders. However, for various reasons, it may not be always possible
(and desirable too) to arrange a physical presence of all the bidders
at the same place. Further, it may not be desired that the bid of
a party gets influenced by the bid of another as it happens in publicly
organized English and Dutch actions, where the bids are public information.
This led to the idea of another class of auction mechanism which is
referred to as sealed-bid second-price auction. In this frequently
used scheme for auction, bids are private information instead of public
announcement. Further, the bids are made simultaneously instead of
one after another, and computation of all the bids is done together,
and the bidder who announces the highest bid, wins the bid, and the
winner bidder pays the price of the second-highest bid. These features
distinguish it from English and Dutch auction schemes. As this type
of auction demands security of the bids (to keep them private) it
reduces to an important cryptographic problem. Several solutions using
classical cryptographic methods have been proposed for this problem
\cite{michiharu1998secure,juels2002two,watanabe2000reducing,suzuki2000efficient,kikuchi1999multi},
but the fact that classical cryptography cannot provide an unconditional
security whereas quantum cryptography can (cf. \cite{pathak2013elements}),
led to a set of resent proposals for quantum sealed-bid auction- where
quantum resources are used to perform sealed-bid auction in an unconditionally
secure manner. The first quantum sealed-bid auction scheme was proposed
by Naseri in 2008 \cite{naseri2009secure}. In 2009, various attack
and defense strategies were proposed in connection with the Naseri
protocol \cite{yang2009improved,qin2009cryptanalysis,zheng2009comment,liu2009revisiting}.
In 2010, Zhao et al., proposed a new sealed-bid auction scheme with
post-confirmation \cite{zhao2010secure}. In the same year, Zhang
proposed another scheme for quantum sealed-bid auction using EPR pairs
\cite{zhang2010quantum}. Subsequently, in 2014, Zhao improved this
post-confirmation-based protocol \cite{zhao2014comment}. The continual
interest on this interesting and upcoming field led to a few more
interesting schemes. For example, in 2015, Wang et al., proposed a
novel quantum sealed-bid auction protocol that utilizes a secret ordering
in the post-confirmation \cite{wang2015new}, and very recently in
2016, Liu et al., have proposed a scheme for multiparty quantum sealed-bid
auction using single photons as message carrier \cite{liu2016multiparty}.
However, all these schemes of quantum sealed-bid auctions have some
limitations, and most of them are vulnerable under specific eavesdropping
strategy. This fact and the importance of the sealed-bid auction schemes
in the modern economy motivated us to design a new protocol for sealed-bid
auction that is free from the limitations of the previously proposed
schemes.

Remaining part of this paper is organized as follows. In Section \ref{sec:Limitations-of-exisiting-schemes},
we discuss the limitations of the existing schemes and thus establish
the motivation for designing a new protocol for quantum sealed-bid
auction that would be free from the limitations of the existing protocols
for the same task. In Section \ref{sec:Proposed-protocol:-CMQKA-based},
we present a new protocol for quantum sealed-bid auction, and illustrate
the working of the same with an explicit example. In Section \ref{sec:Security-analysis},
we critically analyze the security and efficiency of the proposed
scheme. Finally, the paper is concluded in Section \ref{sec:Conclusion}.

\section{Limitations of the existing schemes for quantum sealed-bid auction\label{sec:Limitations-of-exisiting-schemes}}

\begin{figure}
\begin{centering}
\includegraphics[angle=-90,scale=0.4]{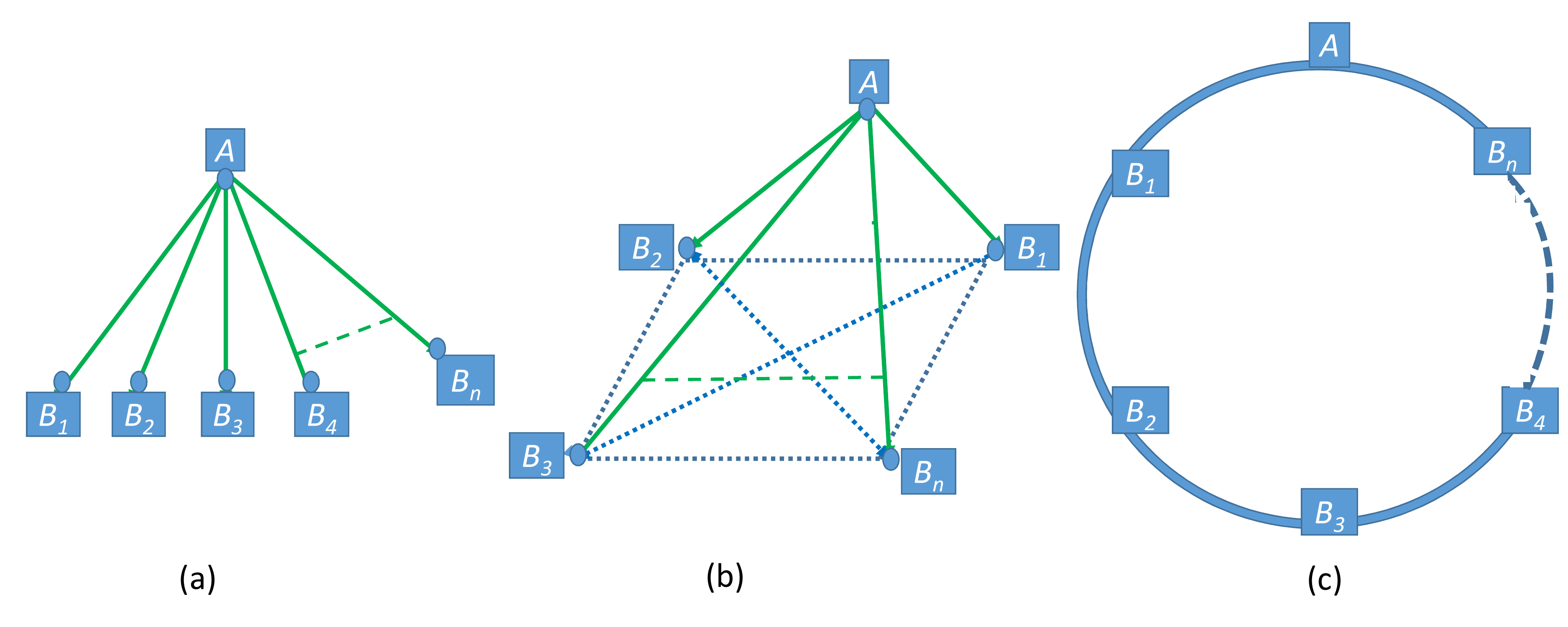}
\par\end{centering}

\protect\caption{\label{fig:Structure}(Color online) Various possible arrangements
of bidders and auctioneer in a (a) tree-type (b) complete-graph-type
(c) circular-type of sealed-bid auction schemes.}
\end{figure}

The limitations of all the existing protocols of quantum sealed-bid
auction may be summarized on the basis of structure of the arrangement
of bidders and the auctioneer in the scheme. Naseri's original protocol
\cite{naseri2009secure} was based on a tree-type structure. In a
tree-type structure (see Fig. \ref{fig:Structure} a), the auctioneer
is like a root while the bidders work as nodes. Every node (bidder)
is required to send his/her information to the root (auctioneer) directly,
and none of the nodes would send his/her information (bid) to the
remaining nodes (bidders). In this type of auction schemes, if the
auctioneer colludes with one of the bidders, he can cheat the auction
process by modifying the colluder's bid, and therefore, announcing
him the winner. In fact, all the remaining bidders have no choice
but to trust the auctioneer.

To circumvent this problem Zhao et al., \cite{zhao2010secure} introduced
a post-confirmation-based technique. All post-confirmation-technique-based
quantum sealed-bid auction schemes convert the tree-type structure
present in the initial schemes of sealed-bid auction to a complete-graph
structure (see Fig. \ref{fig:Structure} b). In the schemes that use
a complete-graph-type structure, each node is directly connected to
all the remaining nodes. Suppose there are $n$ bidders (node) and
an auctioneer then in the post-confirmation-technique-based quantum
sealed-bid auction scheme a complete-graph of $n+1$ node is obtained.
This complete-graph structure has its own limitations related to scalability,
implementation, cost, more traffic, more memory requirement, etc.
One can easily observe that these problems are present in all the
post-confirmation-technique-based quantum sealed-bid auction schemes
\cite{zhao2010secure,zhang2010quantum,zhao2014comment,wang2015new,liu2016multiparty}. 

Further, in all the post-confirmation-technique-based quantum sealed-bid
auction protocols, all the bidders send their bids to each other in
an encoded form, which is used for post-confirmation after the winner
of the auction is announced. They may choose to send the bid values
to each other either before they send it to the auctioneer or after
that. However, both of these kinds of schemes can be attacked. Specifically,
in the former case, the auctioneer may collude with one of the bidders
to modify the bid initially proposed by the colluding bidder, and
thus help him (the colluding bidder) to win the auction. While in
the latter possibility, a single bidder (some of the colluding bidders)
may extract the bid value of another bidder (all other bidders) using
an optimized measurement on the accessible $\left(n-1\right)$ copies
of the bids sent to all the competing bidders. As there exist some
serious attack strategies against complete-graph-type schemes as well
as tree-type schemes, we require a scheme of sealed-bid auction where
all the parties (both auctioneer and bidders) get access to the information
at the same time.

Here, we provide a solution of this problem by proposing a protocol
using circular structure (see Fig. \ref{fig:Structure} c) instead
of tree or complete-graph structures. As there also exist a few collusion
attacks on the circular quantum communication schemes (specifically,
on multiparty circular quantum key agreement schemes), we have incorporated
the solution provided in Refs. \cite{sun2016efficient,sun2016multiparty}
against those attacks on the circular schemes for sending bids to
auctioneer and competing bidders. We will discuss the possible attack
strategies and the relevance of the solution adopted in the security
section. In what follows, we first propose a protocol for sealed-bid
auction that would be free from the limitations of the previous schemes
mentioned in this section. 

\section{Proposed protocol: CMQKA-based scheme for sealed-bid auction\label{sec:Proposed-protocol:-CMQKA-based}}

In this section, we propose a scheme using which the sealed-bid auction
task can be accomplished without confronting the attacks possible
in the schemes proposed in the recent past. Suppose there are $n$
bidders $B_{1},\,B_{2},\,B_{3},\ldots B_{n}$ and an auctioneer. Therefore,
our task may be summed up as each bidder $B_{i}$ wish to send his
bid $b_{i}$ to the auctioneer in a secure manner. Further, he also
needs to share this information with the remaining bidders as well.
At the same time, he must make sure that any single party (or a set
of parties) does not get access to this information until he wishes
them to, i.e., no one should be able to take advantage from this information. 

In our scheme, to initiate the auction all the parties (the auctioneer
and all the bidders) agree on a uniform arrangement in a circular
manner. Additionally, they also decide to use disjoint subgroups (of
a group of unitary operators which provide orthogonal states on the
application) to encode their bid values as discussed in the past for
MQKA scheme \cite{shukla2014protocols}. The disjoint groups are those
groups which have only identity element in common, i.e., $A$ and
$B$ are disjoint groups if $A\cap B=\left\{ I\right\} .$ Requirement
of disjoint subgroups is justified as the receiver will loose the
bijective mapping between each encoding operation and the initial
and final state if two senders would apply the same operation.

Specifically, they obtain $n$ disjoint subgroups of order two, where
one element is identity, to encode each bit of their secret bids.
The bidder $B_{i}$ is assigned a disjoint subgroup $\left\{ U_{I},U_{i}\right\} $,
where $U_{I}$ is the identity operator, and the set of all these
unitary operators $\left\{ U_{I},U_{1},\ldots U_{i},\ldots U_{n}\right\} $
should be a part of a larger group of at least order $n$. It should
be noted that an operational definition of the group (or a modified
group as mentioned in \cite{shukla2014protocols,shukla2013group,banerjee2016asymmetric})
is used hereafter neglecting the global phase.

Before we proceed with the protocol, we would also like to mention
that we have already modified the MQKA scheme \cite{shukla2014protocols}
to circumvent the participant attack mentioned beforehand and in the
security section. Due to this modification, the structural arrangement
of participants in our scheme (which is a circular structure) may
be viewed as an intermediate structure between the tree-type and complete-graph
structures. In what follows, we will establish that tree-type structure
based scheme and complete-graph structure based scheme for quantum
sealed-bid auction can be obtained as special cases of the circular
structure based scheme proposed here. The quantum sealed-bid auction
scheme works as follows, where we assume the bidders and the auctioneer
arranged in a circular manner (cf. Fig. \ref{fig:Structure} (c)).
\begin{description}
\item [{Step1:}] The auctioneer divides all the bidders (arranged in a
large circle) in $l$ number of sub-circles in such a way that each
sub-circle contains equal number of bidders, i.e., $r=\frac{n}{l}+1$.
It should be noted here that only the auctioneer is part of all the
sub-circles. Thus each sub-circle contains $\left(r-1\right)$ distinct
bidders, who are part of only a single sub-circle.\\
Similarly, each bidder also prepares $l$ sub-circles of the remaining
bidders and the auctioneer from the larger circle. The division of
circle in sub-circles by each bidder is similar, but not the same
to each other or auctioneer. In fact, it is important in this arrangement
that the bidder (initiator) himself should remain part of all the
sub-circles he prepares.
\item [{Step2:}] To start the auction, the auctioneer prepares $l$ number
of $\left(r-1\right)$-qubit entangled states $|\psi\rangle^{1},|\psi\rangle^{2},\ldots|\psi\rangle^{l}$,
one for each sub-circle formed in Step1. He may/may not prepare different
entangled states for each sub-circle. Subsequently, he divides each
$\left(r-1\right)$-qubit entangled state into two sequences. The
first sequence containing $p$ number of qubits is called ``travel
qubit sequence $\left(t_{i}\right)$'', and the other one with $\left(r-p-1\right)$
qubits is known as ``home qubit sequence $\left(h_{i}\right)$''. 
\item [{Step3:}] All the bidders also prepare the same number of $\left(r-1\right)$-qubit
entangled states and divide them into two sequences, namely, home
and travel qubit sequences, similar to what the auctioneer has performed
in Step2.
\item [{Step4:}] Auctioneer can begin with the bidding process by sending
his $l$ different travel sequences ($t_{i}$s) to the first bidders
in each sub-circle. Prior sending these qubits, he has to insert an
equal number of decoy qubits randomly in each $t_{i}$ to ensure security
against an adversary \cite{nielsen2010quantum,sharma2016verification}.\\
At the same time all bidders $B_{i}$s also send their $l$ number
of $t_{i}$s (only after inserting an equal number of decoy qubits)
to the first members of their respective $l$ sub-circles. At the
end of this step, all the legitimate parties have received travel
qubits from one of their adjacent party (either auctioneer or a bidder).
Subsequently, they perform a security check using the decoy qubits
inserted in $t_{i}$s among the sender and the receiver of each $t_{i}$.
If they find traces of an eavesdropping attempt inferred from the
errors more than a threshold value, then they abort the protocol,
otherwise proceed to the next step. 
\item [{Step5:}] Once it is ensured that all the parties have received
the $t_{i}$s in a secure manner, they can encode their messages on
it. Specifically, each bidder $B_{i}$ encodes his respective bid
$b_{i}$, and the auctioneer may choose to encode some random bits
or information regarding the auction on the traveling qubits, which
they have received in Step4. For encoding each bit of the secret they
use the prior decided unitary operations $U_{i}$s to obtain the encoded
sequence $t_{i}^{\prime}$.\\
Subsequently, all the parties prepare the same number of decoy qubits
as in the travel sequence and insert them randomly in $t_{i}^{\prime}$
to obtain an enlarged sequence. In fact, at the end of this step,
each bidder and the auctioneer has encoded his information only on
the quantum state prepared by his adjacent party (bidder/auctioneer). 
\item [{Step6:}] The auctioneer and all the bidders send the enlarged sequences
to their next party in their respective sub-circles, i.e., $B_{i}\rightarrow B_{i+1}$.
On the receipt of this sequence each receiving party performs a security
check using the decoy qubits with the sending party following the
same strategy as in Step4. In case of fewer than threshold errors,
each party encodes the bid/information on the received travel sequence
$t_{i}^{\prime}$ using the assigned unitary operations to obtain
$t_{i}^{\prime\prime}$. Finally, he sends this encoded sequence to
the next party only after inserting adequate decoy qubits in it.\\
The same process is repeated for the next $\left(r-3\right)$ steps
until all the $\left(r-1\right)$ parties in each sub-circle have
encoded their information. In the end, the last party in each sub-circle
sends this encoded sequence to the first party who had prepared the
quantum state in a secure manner.
\item [{Step7:}] The auctioneer and all the bidders now possess all the
qubits of $l$ entangled states they have prepared. Therefore, they
can perform a measurement on the entangled particles from the travel
and home qubit sequences in the basis set it was initially prepared.
Due to the presence of a bijective mapping between the set of encoding
operations of all the bidders and auctioneer and the initial-final
entangled state pair, one can easily deduce all the bid values $b_{i}$s.
The auctioneer declares the bidder with the highest bid as the winner
of the auction, which can be easily verified by each bidder. 
\end{description}
In what follows, we will show an explicit example of the proposed
scheme for six bidders and an auctioneer. Before that it is customary
to mention that the proposed scheme has one unexplored advantage that
it can also be performed without an auctioneer. Apart from this, breaking
a larger circle into smaller sub-circles not only provides security
against participants' collusion attack, but also reduces the requirement
of the number of entangled qubits required. As the preparation and
maintenance of a higher dimensional entangled state are difficult,
it becomes significant from the point of view of an experimental realization
of the proposed scheme.

Further, as mentioned beforehand that the proposed scheme can be modified
to obtain both tree-type and complete-graph-type auction schemes (cf.
Fig. \ref{fig:Structure}) as its limiting cases. Specifically, in
our scheme, if $l=n$, i.e., all the bidders are sending their bid
information to the auctioneer and all the remaining bidders, it reduces
to a complete-graph-type of auction scheme. Additionally, if only
the auctioneer prepares the state and receives the bid values from
all the bidders (i.e., the bidders do not send bid values among themselves),
then a tree-type auction scheme can be deduced.

\subsection{An example\label{sub:An-example}}

Here, we show an explicit example in which we suppose there are six
bidders $B_{1},\,B_{2},\,B_{3},\,B_{4},\,B_{5},\,B_{6}$ and one auctioneer
$A$. The bidders ($B_{i}$s) want to encode their bid values $b_{i}$s,
respectively. To initiate the auction all the parties agree on a uniform
arrangement in a circular manner. They further decide their unitary
operations $U_{i}$s to be used for encoding their bids. In this case,
without loss of generality, we may assume that the bidders $B_{2},\,B_{4},$
and $B_{6}$ use the unitary operation $\{I,iY\}$ for encoding their
bids, while $B_{1},B_{3},$ and $B_{5}$ use $\{I$,$X\}$. The auctioneer
encodes using the unitary operations $\{I,Z\}$. It is also predecided
that all the parties will perform $I$ to encode 0, and the other
unitary to encode 1. 

\begin{table}
\centering{}%
\begin{tabular}{|>{\centering}p{1.2cm}|>{\centering}p{1.5cm}|c|c|c|}
\hline 
Parties in auction & Unitary operation $\left(U\right)$ & 1st sub-circle & 2nd sub-circle & 3rd sub-circle\tabularnewline
\hline 
$A$ & $\{I,Z\}$ & $s_{1}^{A}=A,B_{1},B_{2}$ & $s_{2}^{A}=A,B_{3},B_{4}$ & $s_{3}^{A}=A,B_{5},B_{6}$\tabularnewline
\hline 
$B_{1}$ & $\{I$,$X\}$ & $s_{1}^{B_{1}}=B_{1},B_{2},B_{3}$ & $s_{2}^{B_{1}}=B_{1},B_{4},B_{5}$ & $s_{3}^{B_{1}}=B_{1},B_{6},A$\tabularnewline
\hline 
$B_{2}$ & $\{I,iY\}$ & $s_{1}^{B_{2}}=B_{2},B_{3},B_{4}$ & $s_{2}^{B_{2}}=B_{2},B_{5},B_{6}$ & $s_{3}^{B_{2}}=B_{2},A,B_{1}$\tabularnewline
\hline 
$B_{3}$ & $\{I$,$X\}$ & $s_{1}^{B_{3}}=B_{3},B_{4},B_{5}$ & $s_{2}^{B_{3}}=B_{3},B_{6},A$ & $s_{3}^{B_{3}}=B_{3},B_{1},B_{2}$\tabularnewline
\hline 
$B_{4}$ & $\{I,iY\}$ & $s_{1}^{B_{4}}=B_{4},B_{5},B_{6}$ & $s_{2}^{B_{4}}=B_{4},A,B_{1}$ & $s_{3}^{B_{4}}=B_{4},B_{2},B_{3}$\tabularnewline
\hline 
$B_{5}$ & $\{I$,$X\}$ & $s_{1}^{B_{5}}=B_{5},B_{6},A$ & $s_{2}^{B_{5}}=B_{5},B_{1},B_{2}$ & $s_{3}^{B_{5}}=B_{5},B_{3},B_{4}$\tabularnewline
\hline 
$B_{6}$ & $\{I,iY\}$ & $s_{1}^{B_{6}}=B_{6},A,B_{1}$ & $s_{2}^{B_{6}}=B_{6},B_{2},B_{3}$ & $s_{3}^{B_{6}}=B_{6},B_{4},B_{5}$\tabularnewline
\hline 
\end{tabular}\protect\caption{\label{tab: Subcircle Table} The unitary operations assigned to each
party along with the arrangement of sub-circles are mentioned explicitly
across the auctioneer $A$ and each bidder $B_{i}$ in an example
scheme with one auctioneer and six bidders in a circular sealed-bid
auction scheme.}
\end{table}

\begin{figure}
\begin{centering}
\includegraphics[angle=-90,scale=0.5]{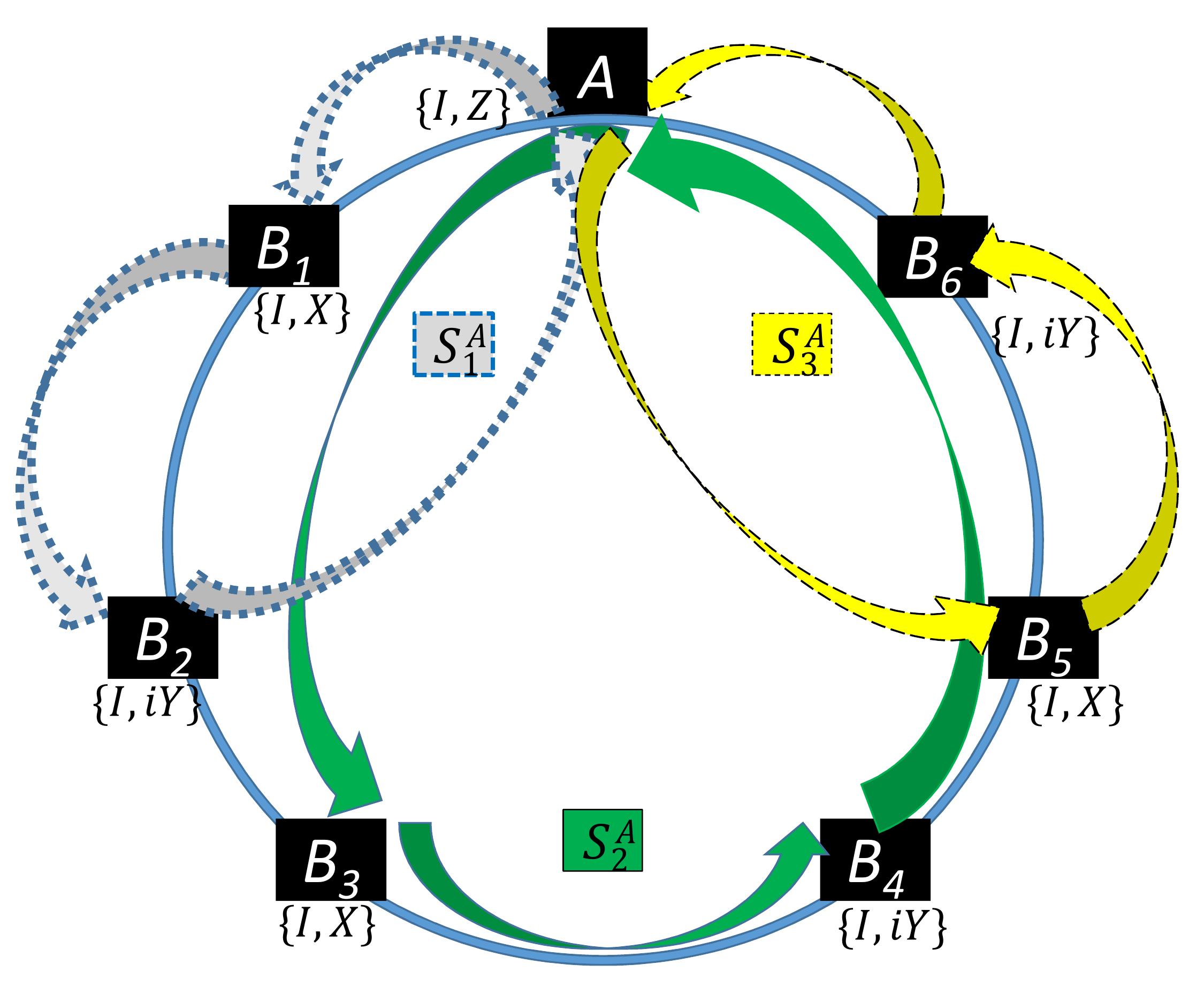}
\par\end{centering}

\protect\caption{\label{fig:Circular-structure-of}(Color online) A circular-type of
sealed-bid auction scheme. The auctioneer $\left(A\right)$ and all
the bidders $\left(B_{i}\right)$ are arranged in a circle shown in
smooth (blue) line. The Sub-circle $s_{1}^{A}$ is presented by gray
(dotted) arrows; Sub-circle $s_{2}^{A}$ by green (solid dense) arrows;
and Sub-circle $s_{3}^{A}$ by yellow (solid semi-transparent) arrows.}
\end{figure}

The working of our example scheme is summarized in the following steps.
\begin{description}
\item [{Example-Step1:}] In this example, we wish to ensure that fewer
than four participants cannot successfully collude to cheat. For the
same the auctioneer (and all bidders) divide all the participants
in $l=3$ number of sub-circles in such a way that each sub-circle
contains an equal number of bidders, i.e., $r=\frac{n}{l}+1=\frac{6}{3}+1=3$.
One such arrangement is shown in Fig. \ref{fig:Circular-structure-of}.
Note that only $A$ is present in all the sub-circles, otherwise all
the bidder are a part of only single sub-circle. Namely, the Sub-circle
$s_{1}^{A}$ contains $A,\,B_{1},\,B_{2}$; Sub-circle $s_{2}^{A}$
contains $A,\,B_{3},\,B_{4}$; and Sub-circle $s_{3}^{A}$ contains
$A,\,B_{5},\,B_{6}$. Using a similar approach all the bidders make
their three sub-circles which are mentioned in detail in Table \ref{tab: Subcircle Table}.
\item [{Example-Step2:}] The auctioneer prepares three distinct 2-qubit
entangled states (Bell states) to receive each bit value of the secret
bids from each bidder. For example, he may prepare three copies of
$|\psi^{+}\rangle=\frac{1}{\sqrt{2}}\left(|00\rangle+|11\rangle\right)$.
He may also to use different entangled state for each sub-circle,
like $|\psi^{+}\rangle$ for $s_{1}^{A}$, $|\psi^{-}\rangle$ for
$s_{2}^{A}$, and $|\phi^{+}\rangle$ for $s_{3}^{A}$. In general,
he prepares $3N$ Bells states independently to receive bids of $N$-bits
from all the bidders. Then he divides all three sets of $N$ Bell
states (prepared for each sub-circle) to obtain two sequences of all
the first qubits $h_{i}$ (home qubit sequence) and the second qubits
$t_{i}$ (travel qubit sequence).
\item [{Example-Step3:}] All the bidders also prepare the same number of
Bell states and obtain two sequences of home and travel qubits as
the auctioneer in previous step. \\
Column 2 of Table \ref{tab: example} lists three Bell states each
party prepares to obtain information regarding $i$th bit of the secret,
like $\left(|\psi_{s_{1}^{J}}\rangle\otimes|\psi_{s_{2}^{J}}\rangle\otimes|\psi_{s_{3}^{J}}\rangle\right)$
is the $i$th set of Bell states prepared by $J$th party.
\item [{Example-Step4:}] The auctioneer sends all three travel qubit sequences
$t_{i}$s to the first bidders (i.e., $B_{1},\,B_{3},\,B_{5}$) of
all three sub-circles $s_{1}^{A},\,s_{2}^{A},\,s_{3}^{A}$, respectively.
He also inserts an equal number of decoy qubits randomly in all the
traveling sequences. At the same time all the bidders $B_{i}$ also
send their $t_{i}$s (after inserting decoy qubits) to the next bidder
$B_{i+1}$ in their respective sub-circles $s_{j}^{B_{i}}$. All the
recipients now perform the security check with corresponding senders
using decoy qubits; and if they find an eavesdropper they abort the
protocol, otherwise continue the process. 
\item [{Example-Step5:}] Each bidder encodes his respective bid $b_{i}$,
and the auctioneer encodes random bits on $t_{i}$, which they had
received in the previous step, using a unitary operation $U$ mentioned
in Column 2 of Table \ref{tab: Subcircle Table}. In fact, Column
3 of Table \ref{tab: example} lists the operations performed in this
round. Though we have not mentioned the transformed state in the table,
it is a trivial task to obtain that the auctioneer's (and similarly
other's) transformed state becomes $\left(U_{B_{1}}|\psi_{s_{1}^{A}}\rangle\otimes U_{B_{3}}|\psi_{s_{2}^{A}}\rangle\otimes U_{B_{5}}|\psi_{s_{3}^{A}}\rangle\right)$. 
\item [{Example-Step6:}] Subsequently, all the parties send the encoded
qubits to the next party in a secure manner, who also performs the
unitary operations corresponding to his bit value before sending the
travel qubits finally to the party who has prepared it. The unitary
operations performed on the quantum state are mentioned in Column
4 of Table \ref{tab: example}. The auctioneer's quantum state may
be written as $\left(U_{B_{2}}U_{B_{1}}|\psi_{s_{1}^{A}}\rangle\otimes U_{B_{4}}U_{B_{3}}|\psi_{s_{2}^{A}}\rangle\otimes U_{B_{6}}U_{B_{5}}|\psi_{s_{3}^{A}}\rangle\right)$.\\
It is interesting to see here that the combined state contains the
encoding from all the bidders. Similarly, all the parties hold the
message encoded travel qubits and home qubits they have prepared.
\item [{Example-Step7:}] Finally, the auctioneer and all the bidder perform
the Bell state measurement on all the three states and extract the
information regarding each $U_{i}$s. Using which a winner may be
chosen.
\end{description}
In Table \ref{tab: example}, it is important to note that the operation
in Round 2 (Column 4) can be obtained from the operation in Round
1 (Column 3), just by performing a cyclic permutation of rows. This
can be attributed to the cyclic property of the scheme proposed here.

\begin{table}
\centering{}%
\begin{tabular}{|>{\centering}p{1.2cm}|c|c|c|}
\hline 
Parties in auction & Initial $i$th set of Bell states & Operations in Round 1 & Operations in Round 2\tabularnewline
\hline 
$A$ & $|\psi_{s_{1}^{A}}\rangle|\psi_{s_{2}^{A}}\rangle|\psi_{s_{3}^{A}}\rangle$ & $\left(U_{B_{1}}\otimes U_{B_{3}}\otimes U_{B_{5}}\right)$ & $\left(U_{B_{2}}\otimes U_{B_{4}}\otimes U_{B_{6}}\right)$\tabularnewline
\hline 
$B_{1}$ & $|\psi_{s_{1}^{B_{1}}}\rangle|\psi_{s_{2}^{B_{1}}}\rangle|\psi_{s_{3}^{B_{1}}}\rangle$ & $\left(U_{B_{2}}\otimes U_{B_{4}}\otimes U_{B_{6}}\right)$ & $\left(U_{B_{3}}\otimes U_{B_{5}}\otimes U_{A}\right)$\tabularnewline
\hline 
$B_{2}$ & $|\psi_{s_{1}^{B_{2}}}\rangle|\psi_{s_{2}^{B_{2}}}\rangle|\psi_{s_{3}^{B_{2}}}\rangle$ & $\left(U_{B_{3}}\otimes U_{B_{5}}\otimes U_{A}\right)$ & $\left(U_{B_{4}}\otimes U_{B_{6}}\otimes U_{B_{1}}\right)$\tabularnewline
\hline 
$B_{3}$ & $|\psi_{s_{1}^{B_{3}}}\rangle|\psi_{s_{2}^{B_{3}}}\rangle|\psi_{s_{3}^{B_{3}}}\rangle$ & $\left(U_{B_{4}}\otimes U_{B_{6}}\otimes U_{B_{1}}\right)$ & $\left(U_{B_{5}}\otimes U_{A}\otimes U_{B_{2}}\right)$\tabularnewline
\hline 
$B_{4}$ & $|\psi_{s_{1}^{B_{4}}}\rangle|\psi_{s_{2}^{B_{4}}}\rangle|\psi_{s_{3}^{B_{4}}}\rangle$ & $\left(U_{B_{5}}\otimes U_{A}\otimes U_{B_{2}}\right)$ & $\left(U_{B_{6}}\otimes U_{B_{1}}\otimes U_{B_{3}}\right)$\tabularnewline
\hline 
$B_{5}$ & $|\psi_{s_{1}^{B_{5}}}\rangle|\psi_{s_{2}^{B_{5}}}\rangle|\psi_{s_{3}^{B_{5}}}\rangle$ & $\left(U_{B_{6}}\otimes U_{B_{1}}\otimes U_{B_{3}}\right)$ & $\left(U_{A}\otimes U_{B_{2}}\otimes U_{B_{4}}\right)$\tabularnewline
\hline 
$B_{6}$ & $|\psi_{s_{1}^{B_{6}}}\rangle|\psi_{s_{2}^{B_{6}}}\rangle|\psi_{s_{3}^{B_{6}}}\rangle$ & $\left(U_{A}\otimes U_{B_{2}}\otimes U_{B_{4}}\right)$ & $\left(U_{B_{1}}\otimes U_{B_{3}}\otimes U_{B_{5}}\right)$\tabularnewline
\hline 
\end{tabular}\protect\caption{\label{tab: example} The scheme is summarized here with the $i$th
triplet of Bell states (one for each sub-circle) prepared by each
party to obtain $i$th bit value of the secret from all the remaining
parties as mentioned in the second column against each party. The
operations performed on the initial state in the first and second
rounds are also explicitly mentioned in the same rows in the last
two columns. Here, $U_{i}=I\otimes U_{e}$, where $U_{e}$ is the
unitary corresponding to the bit value a user wants to encode and
are mentioned in Column 2 of Table \ref{tab: Subcircle Table}.}
\end{table}

\section{Security and efficiency analysis\label{sec:Security-analysis}}

A sealed-bid auction process may be viewed as a multiparty computational
task, where each party sends inputs to the auctioneer, who computes
the value of the function (i.e., maximum/minimum of all the inputs/bids).
The security of a multiparty scheme should be ensured both against
an outsider and insider attacker. In fact, an insider attacker is
more powerful than an outsider attacker. Here, we will show the security
of our sealed-bid auction scheme against both types of attacks.

\subsection{Outsider's attacks}

As the outsider's attacks can be checked by using the standard eavesdropping
checking techniques (subroutines) discussed in Ref. \cite{sharma2016verification}.
Here, we restrict ourselves from elaborating on them. We just note
that it is straight forward to observe that the use of BB84 subroutine
or GV subroutine can protect our scheme at least from the following
types of outsider's attacks: (i) Intercept-resend attack, (ii) Entanglement
measure attack, (iii) Disturbance or modification attack, (iv) Impersonation
or man-in-the-middle attack, and (v) Trojan-horse attack.

Specifically, the first three attacks may be foiled by the use of
decoy qubits inserted by each sender. An eavesdropping attack using
one of these three techniques will always leave detectable traces
at the receiver's end (for details see \cite{banerjee2016asymmetric}).
Further, the disturbance attack is only a type of denial of service
attack so Eve do not get any advantage through it, but she may use
it to interrupt the auction. 

To check the impersonation attack, a suitable authentication protocol
is used and all the parties may inform regarding sending and receiving
of the qubits via an authenticated classical channel. It is also known
that there exist suitable technical measures to circumvent trojan-horse
attack. Keeping this in mind, we now proceed to discuss insider's
attacks in detail.

\subsection{Insider's attacks}
\begin{enumerate}
\item \textbf{Participant attack: }A single bidder may also try to attack
the auction scheme. For example, a bidder $B_{i}$ may choose to send
different bid values to the auctioneer and each voter. As a specific
case, consider that the bid value sent to the auctioneer is higher
than that sent to the remaining bidders, and the auctioneer chooses
him as the winner of the auction and announces his bid value. However,
the tally of the remaining bidders will show different value of the
bid by that bidder. This may be ascribed as a case of cheating and
the auction will be called off. However, if $B_{i}$ wishes to disrupt
the auction process, he may do so by using this strategy. In summary,
a participant will not gain any benefit from this attack as he will
always be detected. Therefore, this attack can be viewed as a denial
of service attack.
\item \textbf{Collusion attack: }Another interesting insider's attack on
a multiparty scheme is a collusion attack. Our scheme being a circular-type
scheme of multiparty quantum communication, increases the number of
parties need to collude to obtain any useful information. To elaborate
this point, we may consider that the bidders $B_{i}$ and $B_{i+m}$
collude to learn the encoding of all the $m$ bidders lying between
them in the circle. Best strategy for them would be that $B_{i}$
would generate an entangled state having an adequate number of qubits,
and would send the same number of qubits that $B_{i}$ has received
from $B_{i-1}$ to $B_{i+1}$. He would send the remaining qubits
of the entangled states directly to $B_{i+m}$. At a later time, $B_{i+m}$
will receive the set of travel qubits form $B_{i+m-1}$, which would
have encoded bidding information of all the $\left(m-1\right)$ bidders'
lying between them. As $B_{i}$ has already sent the home qubits of
the entangled state he had prepared to $B_{i+m}$, he may now use
both set of qubits and the knowledge of initial state prepared by
$B_{i}$ to obtain information of all the $\left(m-1\right)$ bidders.\\
However, this attack will not be as effective as they wish it to be.
Specifically, an effective attack that can deterministically affect
the outcome of the bidding process would require that $B_{i}$ and
$B_{i+\frac{n}{2}}$ collude, as in that case, the colluding parties
would obtain the bid values of all the other bidders. Precisely, both
$B_{i}$ and $B_{i+\frac{n}{2}}$ prepare entangled states and send
their travel qubits to $B_{i+1}$ and $B_{i+\frac{n}{2}+1}$ and home
qubits to $B_{i+\frac{n}{2}}$ and $B_{i}$, respectively. After $n/2$
rounds, when both $B_{i}$ and $B_{i+\frac{n}{2}}$ receive the encoding
of all the remaining party they can choose their appropriate bid values.
Further, it is important to note that at least one of the colluding
parties will always have access to the travel qubits after they have
learnt other's secret. Therefore, they will leave no traces in this
kind of an attack. \\
Here, the security against this attack is achieved here by breaking
the circle into $l$ sub-circles. In this case, if less than $l$
attackers collude, then they can not cheat the remaining bidders and
the auctioneer. Thus, increase in the number $l$ would decrease the
effectivity of this attack.
\end{enumerate}
There is a quantitative measure to analyze the efficiency of a quantum
communication scheme known as the qubit efficiency, proposed in Ref.
\cite{cabello2000quantum}, given by $\eta=\frac{c}{q+b}$. Here,
$c$ corresponds to the number of classical bits transmitted with
the help of $q$ number of qubits and $b$-bits of classical communication.
We will show the efficiency of the example protocol to give an idea
about the performance of the proposed scheme. However, qubit efficiency
of the proposed scheme can also be calculated using the same approach.

In the example of the proposed scheme qubit efficiency for sending
each bit of bids from all the bidders can be calculated as follows.
In Sub-circle $s_{1}^{A}$, $c_{A}=2$ bits of the message regarding
bids of $B_{1}$ and $B_{2}$ were encoded using a Bell state ($q_{A}^{\prime}=2$).
Additionally, one decoy qubit was inserted by each party in the sub-circle
(i.e., $A,\,B_{1},$ and $B_{2}$) resulting in total $d_{A}=3$.
Therefore, total number of qubits used are $q_{A}=q_{A}^{\prime}+d_{A}=2+3=5$.
Further, no classical communication is involved so $b_{A}=0$. The
same scenario is repeated in Sub-circles $s_{2}^{A}$ and $s_{3}^{A}$.
Hence, for all the sub-circles initiated by $A$, the total contribution
in $c$ and $q$ are thrice of that calculated for one sub-circle,
i.e., 6 and 15, respectively. 

In one of the sub-circles initiated by the bidders $B_{i}$, the auctioneer
$A$ do not encode a useful message so it is not counted as classical
bit. Therefore, in six sub-circles $\left(s_{3}^{B_{1}},\,s_{3}^{B_{2}},\,s_{2}^{B_{3}},\,s_{2}^{B_{4}},\,s_{1}^{B_{5}},\,s_{1}^{B_{6}}\right)$,
where $A$ is present contribution in $c$ becomes 1 bit, while the
number of qubits used remain the same. The remaining 12 sub-circles
initiated by the bidders will have the same contribution in the calculations
of $c$ and $q$ as that of the sub-circle initiated by $A$. Therefore,
the total amount of classical information transmitted is $c=2\times\left(3+12\right)+1\times6=36$
bits and the same is done using $q=5\times21=105$ qubits, and involving
no classical communication $b=0$ (except for eavesdropping checking).
Finally, the auctioneer announces the winner out of six competing
bidders, which corresponds to $b=\log_{2}6=2.585$ bits of additional
classical communication. Hence, the qubit efficiency of the proposed
scheme (with each bid of $n$ bits) turns out to be $\eta=\frac{36n}{105n+\log_{2}6}\approx34.28\%$
(if $n\gg2.585)$. In what follows, we compare the qubit efficiency
of various schemes, and to do so, we have considered $n\gg2.585$
to be consistent. 

The qubit efficiency of a tree-type sealed-bid auction (without post-confirmation)
scheme \cite{yang2009improved} that would do a task similar to what
is done in the proposed scheme (however, at the end of the bidding
process none of the bidders would obtain the bid of others as it happens
in our scheme) is calculated as $\frac{6n}{20n+\log_{2}6}\approx30\%$,
which appears to be lower than the qubit efficiency of the proposed
scheme.\textcolor{red}{{} }On the contrary, a tree-type scheme would
appear more efficient in comparison with the proposed scheme, if we
consider that the task (auction process) is restricted to a situation
where bidding information is only communicated to the auctioneer.\textcolor{red}{{}
}To be specific, we may consider that the information communicated
by each bidder to all other bidders is not meaningful information
(as far as the task is concerned), and it does not contribute to $c$,
which corresponds to meaningful classical information. In that case,
the efficiency of the proposed scheme would be $\frac{6n}{105n+\log_{2}6}\approx5.71\%$.\textcolor{red}{{}
}In the above discussed situation, qubit efficiency of a sealed-bid
auction scheme based on complete-graph structure and proposed to be
realized using single photons \cite{liu2016multiparty} is obtained
as $\frac{6n}{83n+\log_{2}6}\approx7.23\%$, which should be modified
to $\frac{6n}{89n+\log_{2}6}\approx6.74\%$ after concatenating some
decoy photons for security as mentioned in \cite{wang2014revisiting}.
Similarly, the efficiency of the sealed-bid auction scheme using 7-qubit
GHZ states \cite{luo2013loophole} is calculated as $\frac{6n}{92n+\log_{2}6}\approx6.52\%$,
which falls to $\frac{6n}{98n+\log_{2}6}\approx6.12\%$ due to the
addition of some decoy photons in accordance with the scheme proposed
in Ref. \cite{wang2014revisiting}.

There is an interesting feature of the proposed scheme- it can be
performed without the auctioneer. This is so because, each party obtains
bidding information of the others in the post-confirmation stage.
If bidding is performed without an auctioneer in that case the qubit
efficiency of the scheme would have been $\frac{6n}{90n}=6.67\%$,
which is comparable with the qubit efficiency of the schemes that
use complete-graph structure. In brief, our scheme is only slightly
less efficient compared to the schemes based on complete-graph structure,
but it provides higher security against collusion attacks and requires
quantum resources that are easier to prepare.\textcolor{red}{{} }

\section{Conclusion\label{sec:Conclusion}}

We proposed a scheme for quantum sealed-bid auction, which is free
from the limitations of the existing schemes for the same purpose
proposed by other groups, and have shown that the proposed scheme
is not only secure it is also efficient. The advantages of the present
scheme are actually obtained by transforming a complete-graph structure
to a circular structure and subsequently transforming that to sub-circles.
This unique strategy and the benefits obtained by adopting this strategy
are expected to open up a new window for more research related to
quantum auction for a couple of reasons. Firstly, sealed-bid auction
is an extremely important process in our daily life, secondly, with
an increase in $l$ (the number of sub-circles) the size of the entangled
state required reduces whereas the security against collusion attack
increases. This trade-off has a fantastic effect as at present preparation
and maintenance of $s$-qubit entangled state is very difficult when
$s$ is large. However, in our case smaller $s$ leads to better security
in one hand and scalability on the other. It seems feasible that the
proposed scheme can be realized experimentally using available technologies,
but this was not the case with most of the other proposals as $s$
was very high for them. Further, auction process demands security
and quantum version of the same can provide unconditional security
which is not possible by any of its classical counterparts. Keeping
this in mind, we conclude this paper with an expectation that the
works reported here will be realized experimentally and will find
applications in daily life.

\textbf{Acknowledgment: }AP and KT thank Defense Research \& Development
Organization (DRDO), India for the support provided through the project
number ERIP/ER/1403163/M/01/1603. 

\bibliographystyle{plain}
\bibliography{reference-sealed-bid-auction}

\end{document}